\documentclass{article}

\PassOptionsToPackage{numbers, compress}{natbib}

\usepackage[dblblindworkshop, final]{neurips_2026}
\workshoptitle{Foundations of Language Model Security: Theory, Practice and Fundamental Limits}

\usepackage[utf8]{inputenc}
\usepackage[T1]{fontenc}
\usepackage{hyperref}
\usepackage{url}
\usepackage{booktabs}
\usepackage{amsfonts}
\usepackage{amsmath}
\usepackage{nicefrac}
\usepackage{microtype}
\usepackage{xcolor}
\usepackage{amsthm}
\usepackage{graphicx}
\newtheorem{theorem}{Theorem}
\newtheorem{proposition}{Proposition}
\newtheorem{corollary}{Corollary}

\title{MIRROR: Multipath Quorum Integrity for {LLM} Multi-Agent Communication}

\author{%
  Ryuichi Yamafuji Lun \\ University of Southern California \\ \texttt{rlun@usc.edu} \\
  \And Jingzhen Wang \\ University of Southern California \\ \texttt{jwang977@usc.edu} \\
  \AND Shreyas Kolte \\ University of Southern California \\ \texttt{kolte@usc.edu} \\
  \And Ruiteng Li \\ University of Southern California \\ \texttt{ruitengl@usc.edu} \\
}

\begin{document}

\maketitle

\begin{abstract}
Inter-agent communication is central to Large Language Model Multi-Agent Systems (LLM-MAS), but it introduces an underexplored vulnerability: Agent-in-the-Middle (AiTM) attacks that manipulate messages in transit without compromising the agents themselves. Prior work reports Attack Success Rates (ASR) approaching 100\% on structured tasks. Existing defenses rely on semantic validation, which requires additional inference and can block benign outputs, or on transport-layer encryption, which does not help when an intermediary legitimately terminates TLS. We present MIRROR, a communication-layer integrity primitive that replicates a single canonicalized payload across $k$ logical routes and accepts a message only when a strict majority of routes report the same digest. We are explicit about where the security comes from. MIRROR uses \emph{unkeyed} hashing and so authenticates nothing on its own, since an active on-path adversary can always recompute a digest over a payload it has modified. All integrity derives from the assumption that honest routes form a majority. The digest serves only to make witness routes constant-size and to bind the recovered payload to the quorum-agreed value under second-preimage resistance. We state this scope precisely, give the guarantee under a route-compromise bound $\alpha < 0.5$, and extend it to \emph{correlated} routes, where the quantity that matters is the size of the largest shared-failure group and not the route count. We further show that availability and integrity degrade at the same threshold: below $\alpha = 0.5$, quorum-denial and message-dropping adversaries cannot block honest traffic. Across MMLU, HumanEval, and MBPP on two frameworks and four communication topologies, and in a MetaGPT deployment against a production API, MIRROR reduces ASR to 0\% below the threshold at $1\times$ LLM token cost. LLM-as-a-Judge costs $35\times$ in the same deployment, and blocks up to 44.2\% of benign outputs in the topology sweep.
\end{abstract}

\section{Introduction}

The rapid adoption of LLMs has shifted applications from single-model prompting to multi-agent systems (LLM-MAS), where agents collaborate via natural language \citep{wang2024survey, xi2023rise}. Frameworks such as AutoGen \citep{wu2023autogen}, CAMEL \citep{li2023camel}, and MetaGPT \citep{hong2023metagpt} exemplify this paradigm. As these systems move toward deployment across organizational boundaries, the integrity of inter-agent communication becomes critical. The communication layer nonetheless remains largely unprotected: Agent-in-the-Middle (AiTM) attacks manipulate messages in transit and corrupt downstream reasoning, reaching 100\% ASR on code-generation tasks in an undefended AutoGen chain (Appendix~\ref{app:baseline}).

\paragraph{Why transport encryption is not sufficient.}
A natural objection is that TLS already prevents message tampering. It does, for a network-level eavesdropper. It does not help in the deployment we target, where agents communicate through third-party API gateways or broker services that \emph{legitimately terminate TLS} and therefore see and can modify the plaintext payload. End-to-end digital signatures would address this, but require provisioning, storing, and rotating private keys inside agent frameworks. That is an operational burden across decentralized deployments, and a poor fit for systems whose components are prone to leaking context. MIRROR targets the gap between these two: stronger than TLS against a terminating intermediary, and deployable without a PKI.

\paragraph{Scope and framing.}
We deliberately avoid the language of Byzantine fault tolerance. Classical BFT \citep{lamport1982byzantine, castro1999pbft} concerns node-level faults, multi-round consensus, and state-machine replication; MIRROR concerns untrusted \emph{intermediaries} on a single-shot message path, and provides no agreement over agent outputs and no protection against a compromised agent. We use the term \emph{multipath quorum integrity} and contrast the two settings in Table~\ref{tab:bft_contrast}. We claim no algorithmic novelty relative to the distributed-systems literature on quorum-based integrity; the contribution is a precise adaptation, scoping, and evaluation of that mechanism for LLM-MAS, together with an analysis of the route-independence assumption on which it actually rests.

\paragraph{Contributions.}
\begin{itemize}
    \item \textbf{A precisely scoped communication-layer primitive.} We specify MIRROR's mechanism, adversary capabilities, endpoint assumptions, and, critically, what the unkeyed digest does and does not provide (Section~\ref{sec:mechanism}).
    \item \textbf{Guarantees under correlated routes.} We show that the security threshold is governed by the size of the largest shared-failure group, which yields a concrete deployment rule (Proposition~\ref{prop:correlated}).
    \item \textbf{Availability is not traded away.} Below the compromise threshold, quorum-denial and drop adversaries cannot block honest traffic (Corollary~\ref{cor:availability}).
    \item \textbf{Empirical validation} on AutoGen and CAMEL across four communication topologies, and on MetaGPT against a production API (Section~\ref{sec:eval}).
\end{itemize}

\section{System and threat model}
\label{sec:threat}

\paragraph{System model.}
We consider an LLM-MAS with agents $\mathcal{A} = \{A_1, \dots, A_n\}$ exchanging natural-language messages. At each step an agent receives incoming messages and produces a response forwarded to others. Each message is an opaque payload $m \in \mathcal{M}$; MIRROR never interprets its contents.

\paragraph{What a route is.}
A \emph{logical route} is one delivery path for a message from sender $S$ to receiver $R$, instantiated as a distinct API gateway, broker service, network path, or region. Routes are not hardware source-routed paths; they are application-level channels chosen by the sending adapter. A message sent over $k$ routes is transmitted $k$ times over $k$ separately addressed channels.

\paragraph{Adversary capabilities.}
We adopt the AiTM model \citep{he2025redteaming} and make the capabilities explicit. On a route it controls, the adversary may \emph{read}, \emph{modify}, \emph{drop}, \emph{delay}, and \emph{replay} messages, and may recompute any unkeyed digest to be consistent with a payload it has substituted. It may coordinate across the routes it controls and knows the protocol, the canonicalization rule, and the hash function. It may not read or modify traffic on routes it does not control, and it cannot break second-preimage resistance of the hash.

\paragraph{Endpoint assumptions.}
MIRROR assumes \emph{honest endpoints}. If the sending or receiving agent is compromised, MIRROR provides no protection: a compromised sender canonicalizes and replicates whatever payload it chooses, across every route. MIRROR is therefore orthogonal to, and composes with, agent-level defenses against prompt injection; it does not replace them.

\paragraph{Compromise bound.}
The adversary controls at most $c$ of $k$ routes with $\alpha = c/k < 0.5$. Section~\ref{sec:correlated} replaces this with a weaker and more realistic condition when routes share infrastructure.

\paragraph{Freshness.}
Payload integrity alone does not prevent replay or reordering. Including a monotonic sequence number and timestamp in the canonicalized byte string extends the same quorum logic to freshness without modification, since replayed messages then fail to match the current-round quorum. We treat this as part of the canonicalization rule.

\section{MIRROR: multipath quorum integrity}
\label{sec:mechanism}

\begin{figure}[t]
\centering
\includegraphics[width=0.72\linewidth, trim=0 120 0 30, clip]{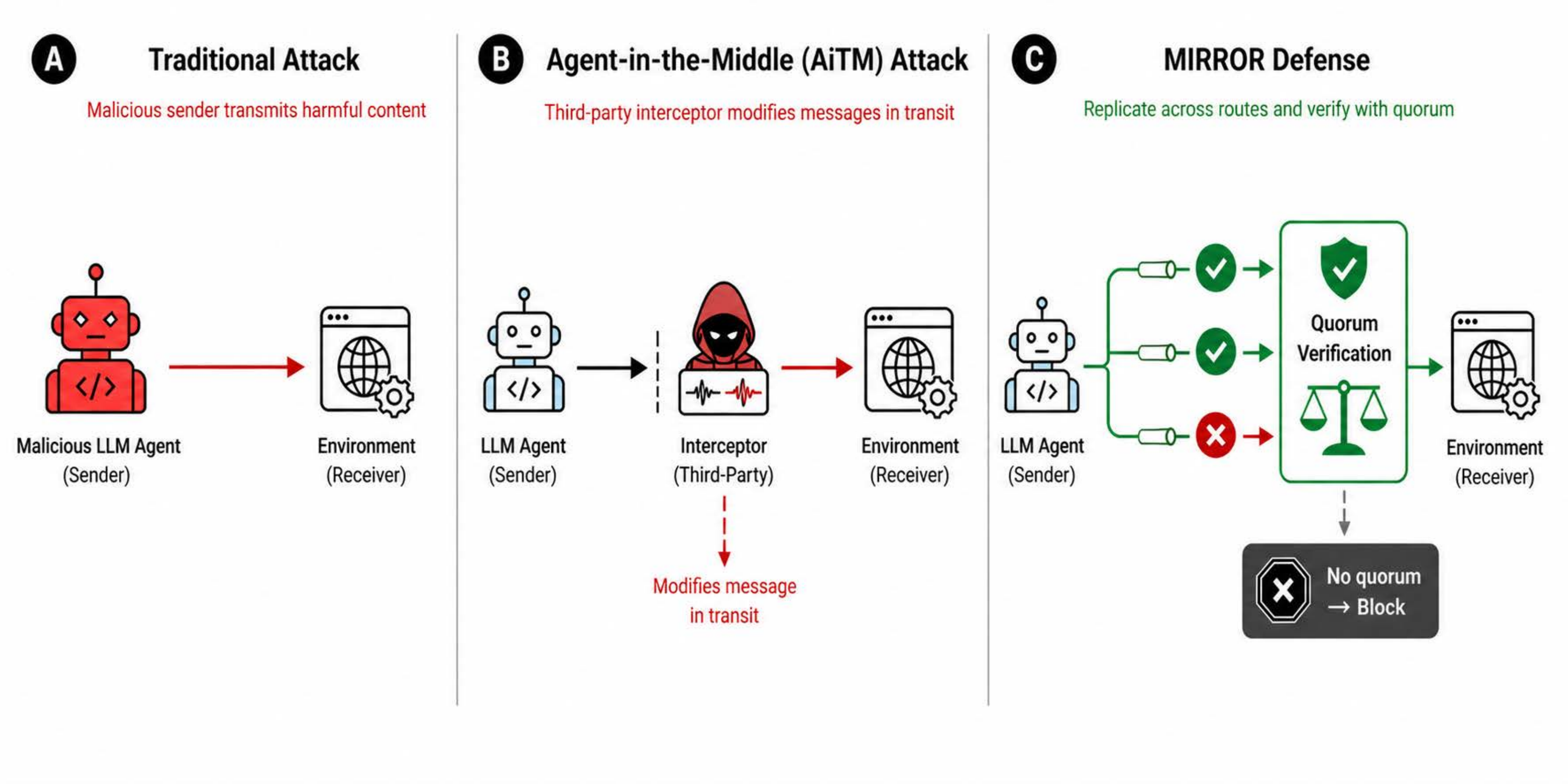}
\caption{Overview of communication threats and the MIRROR defense. (A) Traditional attacks originate from a malicious or compromised sender, which is outside MIRROR's scope. (B) AiTM attacks manipulate messages in transit via a third-party interceptor. (C) MIRROR replicates across routes and verifies quorum, recovering valid messages or blocking corrupted transmissions.}
\label{fig:mirror_overview}
\end{figure}

\subsection{Mechanism}

The sending agent produces \emph{one} response. MIRROR canonicalizes that single finalized output into a byte string, computes its digest, and replicates the identical bytes across routes. Let $k = 2t+1$ be odd. MIRROR assigns $t+1$ routes as \emph{full-payload routes} transmitting $(m, H(m))$, and $t$ routes as \emph{witness routes} transmitting only $H(m)$. We use SHA-256.

The receiver aggregates the reported digests over all $k$ route slots, treating a route that does not deliver within a bounded timeout as reporting no digest. Let $C(H^\ast)$ be the frequency of the most common digest. If $C(H^\ast) > k/2$, the receiver recovers $m$ from any full-payload route whose digest equals $H^\ast$; otherwise it rejects the message. Because the quorum is counted against $k$ rather than against the number of replies received, a delayed or withheld route cannot shift the threshold. Routes deviating from the majority are flagged.

Note that MIRROR queries the LLM \emph{once} per message. It does not require deterministic decoding: the agent may sample at any temperature, because replication happens after generation, over fixed bytes. Deterministic decoding appears in our experiments only as an evaluation control (Section~\ref{sec:eval}).

Our implementation also retires routes that disagree with the quorum, replacing them on subsequent messages. This plays no part in Theorem~\ref{thm:main}, which is per-message and single-round, and we make no claims about it; see Section~\ref{sec:eval} for its role in our runs.

\subsection{What the digest does and does not provide}
\label{sec:hash_scope}

The digest is not a message authentication code. It is unkeyed, carries no secret, and authenticates nothing by itself. An active on-path adversary that modifies a payload simply recomputes the digest over the modified bytes; nothing detects this locally.

Security comes entirely from the quorum. The adversary can make its own routes internally consistent, but doing so only makes them disagree with the honest majority, which costs it the vote. The digest plays exactly two roles:

\begin{enumerate}
    \item \textbf{Compression.} Witness routes carry a fixed 32-byte digest rather than the full payload, so the bandwidth cost of replication is $t+1$ payloads plus $t$ constant-size witnesses rather than $k$ payloads.
    \item \textbf{Binding under second-preimage resistance.} Once $H^\ast$ is fixed by the honest majority, an adversarial full-payload route cannot supply some $m' \neq m$ that the receiver would accept, because that requires $H(m') = H(m)$.
\end{enumerate}

The second role is the only cryptographic assumption we need, and it is \emph{second-preimage} resistance, not collision resistance: the adversary must match a specific target digest it does not choose, not find any colliding pair. We state the assumption accordingly.

\subsection{Guarantee}

\begin{theorem}[Integrity and recovery]
\label{thm:main}
Let $k = 2t+1$ routes carry a message under the scheme above, with $c$ routes adversarial, and assume (i) honest routes transmit identical canonicalized bytes and digests, (ii) $H$ is second-preimage resistant, (iii) endpoints are honest, and (iv) $c < k/2$. Then:
\begin{enumerate}
    \item \textbf{Integrity.} No tampered payload $m' \neq m$ satisfies the quorum condition and is accepted, except with negligible probability.
    \item \textbf{Recovery.} If quorum is reached, the receiver recovers $m$ from a full-payload route matching $H^\ast$, with no additional LLM query.
\end{enumerate}
If $c \geq k/2$ the adversary may control the quorum and neither property holds.
\end{theorem}

\paragraph{Proof sketch.}
Since $c < k/2$, honest routes number $k - c > k/2$ and by (i) all report $H(m)$, so $H^\ast = H(m)$ uniquely. Acceptance of some $m' \neq m$ then requires either a non-$H(m)$ digest reaching frequency $>k/2$, impossible with only $c < k/2$ adversarial routes; or an adversarial full-payload route supplying $m'$ with $H(m') = H^\ast$, a second preimage for a target the adversary does not choose, ruled out by (ii). Recovery follows because $c \le t$ leaves at least one of the $t+1$ full-payload routes honest. The full argument is in Appendix~\ref{app:proof}.

Only the second acceptance path touches cryptography, and it needs second-preimage rather than collision resistance; the remainder of the argument is counting. The substance of the result therefore lies in assumptions (i), (iii), and (iv), which we examine next.

\subsection{Correlated routes}
\label{sec:correlated}

Assumption (iv) treats routes as independently compromisable. In practice, $k$ routes often share a provider, region, or administrative domain, and compromising that shared dependency compromises all of them simultaneously. Because the entire guarantee rests on this assumption, the number of routes is not the appropriate quantity to reason about.

Partition the $k$ routes into $g$ \emph{shared-failure groups} $G_1, \dots, G_g$, where compromising any group's common dependency compromises every route in it. An adversary that compromises a set $\mathcal{S}$ of groups controls $\sum_{j \in \mathcal{S}} |G_j|$ routes.

\begin{proposition}[Threshold under correlation]
\label{prop:correlated}
Theorem~\ref{thm:main} holds against an adversary compromising any group set $\mathcal{S}$ if and only if $\sum_{j \in \mathcal{S}} |G_j| < k/2$. Against a single-group adversary this reduces to $\max_j |G_j| < k/2$.
\end{proposition}

This yields a concrete deployment rule: \emph{no single provider, region, or administrative domain may carry a majority of routes}. For example, provisioning $k=5$ routes through a single provider yields $g=1$ and $\max_j |G_j| = 5$, so no guarantee holds regardless of how large $k$ is. Distributing the same five routes across three providers as $(2,2,1)$ tolerates the compromise of any single provider. The condition can therefore be verified at configuration time; it does not need to be discovered during an attack.

This restates the assumption; it does not remove it. Group assignments are only as reliable as the deployer's knowledge of shared dependencies, and hidden commonality (e.g., two nominally distinct providers reselling the same underlying infrastructure) collapses groups without warning. Measuring route correlation in real deployments therefore remains an open problem, and our evaluation does not address it; our routes are logically instantiated over a shared backend, which corresponds to $g=1$ and exercises the mechanism without testing the assumption.

\subsection{Availability}

A natural concern is that quorum-blocking hands the adversary a denial-of-service vector: deny quorum, and the system halts. Below the compromise threshold this does not occur.

\begin{corollary}[Availability below threshold]
\label{cor:availability}
Let $c_{\mathrm{eff}} = c + f$, where $c$ is the number of adversarial routes and $f$ the number that fail independently. If $c_{\mathrm{eff}} < k/2$, then $k - c_{\mathrm{eff}} > k/2$ honest routes deliver identical digests, so quorum is reached and the honest message is accepted. Injecting distinct digests, dropping, or delaying on adversarial routes cannot prevent this.
\end{corollary}

Integrity and availability therefore fail at the same threshold, and do not trade off against each other: below $\alpha = 0.5$ both hold, above it both are lost. The caveat is that the relevant budget is $c_{\mathrm{eff}}$ and not $c$, so ordinary infrastructure flakiness consumes the same margin as attack. Provisioning $k$ must account for both.

\subsection{Relationship to classical BFT}

\begin{table}[t]
\centering
\small
\setlength{\tabcolsep}{6pt}
\begin{tabular}{lll}
\toprule
 & \textbf{Classical BFT} & \textbf{MIRROR} \\
\midrule
Faulty component & Replica nodes & Network intermediaries \\
Agreement over & Program state / outputs & Message bytes \\
Rounds & Multi-round & Single-shot \\
Requires determinism of & Replica execution & Nothing (post-generation) \\
Protects against bad endpoint & Yes ($<n/3$) & No \\
Threshold & $f < n/3$ (asynchronous) & $c < k/2$ \\
Cryptographic assumption & Signatures / PKI & Second-preimage resistance \\
\bottomrule
\end{tabular}
\caption{MIRROR is not a consensus protocol. It provides integrity for a message in transit under honest endpoints, not agreement among mutually distrusting replicas.}
\label{tab:bft_contrast}
\end{table}

Table~\ref{tab:bft_contrast} makes the boundary explicit. The weaker threshold ($c<k/2$ rather than $f<n/3$) reflects the weaker problem: MIRROR need not tolerate equivocating endpoints, only tampering intermediaries.

\section{Experimental evaluation}
\label{sec:eval}

We ask: \textbf{RQ1:} Does MIRROR prevent AiTM attacks below the threshold, and how does behavior change as adversarial control grows? \textbf{RQ2:} Does it hold across frameworks and topologies? \textbf{RQ3:} What overhead does it impose in a production-API deployment relative to semantic defenses?

\subsection{Setup}

\begin{figure}[htbp]
\centering
\includegraphics[width=\linewidth]{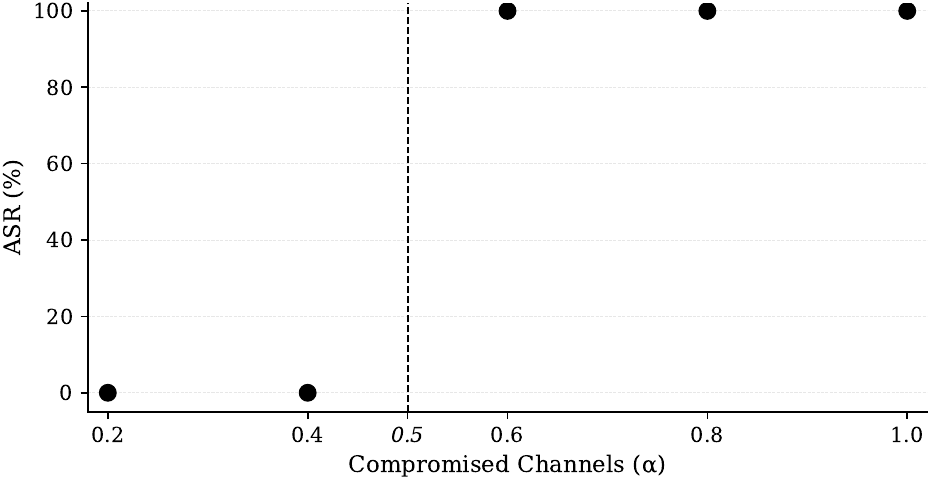}
\caption{ASR on MBPP (AutoGen, $k=5$) as a function of compromised fraction $\alpha = c/k$, for $c \in \{1,\dots,5\}$. The dashed line marks the theoretical threshold $\alpha = 0.5$.}
\label{fig:alpha_sweep}
\end{figure}

We evaluate on \textbf{AutoGen} \citep{wu2023autogen}, \textbf{CAMEL} \citep{li2023camel}, and \textbf{MetaGPT} \citep{hong2023metagpt} with default configurations; on \textbf{MMLU} \citep{hendrycks2020mmlu}, \textbf{HumanEval} \citep{chen2021humaneval}, and \textbf{MBPP} \citep{austin2021program}; using \textbf{Gemma-4-31B-it} \citep{gemma4_2026} locally and \textbf{Gemini 2.5 Pro} \citep{comanici2025gemini25pushingfrontier} via API. RQ1 uses AutoGen and RQ3 uses MetaGPT, both on a 3-agent Chain topology as a worst case for sequential propagation; RQ2 sweeps Chain, Tree, Complete, and Random on AutoGen and CAMEL. Every MIRROR configuration reported uses $k=5$ with SHA-256; the undefended baseline of Appendix~\ref{app:baseline} is $k=1$ by definition.

\textbf{Evaluation controls.} Decoding is deterministic (temperature $0.0$, top-$p$ $1.0$, 1024 max tokens) and route selection uses fixed seeds. As noted in Section~\ref{sec:mechanism}, MIRROR does not require this; it removes generation variance so that measured ASR reflects communication-layer manipulation alone. Routes are instantiated as parallel isolated API requests sharing one backend, a single failure group in the sense of Proposition~\ref{prop:correlated}. We report exhaustive single-pass results over full splits: 164 HumanEval tasks, 427 MBPP test tasks, 707 MMLU STEM examples.

\textbf{Adversary and rotation.} Runs use a \emph{static} adversary, whose compromised set is fixed per run and never re-acquired, with route rotation enabled. A compromised route therefore leaves the adversary's control after its first detected tampering. This does not affect the reported figures, since below the threshold an honest majority yields correct recovery on every message independently of rotation; the 0\% ASR is attributable to quorum verification alone. It does mean we do not test rotation against the adaptive adversary it targets, and we claim no benefit from it.

\textbf{Attack.} No official AiTM implementation is public, so we reimplement from the published description \citep{he2025redteaming}, including the reflection-based adversarial agent and task-specific objectives (Appendix~\ref{app:attack}). The adversary rewrites intercepted messages deterministically and emits \emph{identical} payloads across all compromised routes. This is the worst case for a quorum defense, since vote-splitting would only weaken it. Goals are label-shift on MMLU ($+4$ positions) and injection of a malicious \texttt{safety\_check()} function on code tasks. Because the baseline ASR depends on our reimplementation, we treat absolute baseline magnitudes as indicative and rely on relative comparisons.

\textbf{Metrics.} Attack Success Rate (ASR), the fraction of tasks whose final output is incorrect due to manipulation; and relative token consumption. Compute: one NVIDIA A100 80GB; roughly one hour per API configuration.

\subsection{Behavior across the threshold (RQ1)}

Figure~\ref{fig:alpha_sweep} shows ASR against $\alpha$. For $\alpha < 0.5$ MIRROR achieves 0\% ASR; at $\alpha \geq 0.5$ it transitions sharply to 100\%. This step function is the predicted behavior of Theorem~\ref{thm:main}, not an independent discovery: given the assumptions, the outcome follows by construction. Its value is as a check that the implementation realizes the stated mechanism, and that the transition sits exactly at the predicted threshold instead of degrading earlier, which would indicate canonicalization or routing bugs.

\subsection{Frameworks and topologies (RQ2)}

\begin{table}[htbp]
\centering
\small
\setlength{\tabcolsep}{5pt}
\renewcommand{\arraystretch}{1.05}
\begin{tabular}{l l c c c c}
\toprule
\textbf{Dataset} & \textbf{Framework} & \textbf{No Def.} & \textbf{Judge} & \textbf{FPR (Judge)} & \textbf{MIRROR (k=5)} \\
\midrule
MMLU        & AutoGen & $60.0 \pm 24.9$ & $0.4$ & $0.0$ & $\mathbf{0.0}$ \\
            & CAMEL   & $30.2 \pm 35.8$ & $0.0$ & $10.2$ & $\mathbf{0.0}$ \\
\midrule
MBPP        & AutoGen & $100.0$         & $0.0$ & $0.0$ & $\mathbf{0.0}$ \\
            & CAMEL   & $27.5 \pm 15.5$ & $0.0$ & $44.2$ & $\mathbf{0.0}$ \\
\midrule
HumanEval   & AutoGen & $100.0$         & $0.0$ & $0.0$ & $\mathbf{0.0}$ \\
            & CAMEL   & $37.0 \pm 13.9$ & $0.0$ & $40.9$ & $\mathbf{0.0}$ \\
\bottomrule
\end{tabular}
\caption{ASR (\%) for AutoGen and CAMEL across four topologies (Chain, Tree, Complete, Random), reported as mean $\pm$ standard deviation over the four topologies of Table~\ref{tab:topology_breakdown}. Baselines vary widely by topology (MMLU/CAMEL spans $1.4$--$89.8$); MIRROR is 0.0\% in every configuration. LLM-as-a-Judge also lowers ASR but with substantial false positives.}
\label{tab:main_results}
\end{table}

Without protection, both frameworks are vulnerable, but to sharply differing degrees: AutoGen reaches 100\% ASR on both code-generation benchmarks in every topology, whereas CAMEL ranges from $1.4\%$ to $89.8\%$ on MMLU depending on structure. More informative than the magnitude is that this variation is non-monotonic in topology, so robustness cannot be inferred from communication structure. That is an argument for a structure-independent mechanism.

LLM-as-a-Judge lowers ASR to near zero but at false-positive rates up to 44.2\%: its low ASR partly reflects conservative over-blocking rather than faithful recovery. MIRROR achieves 0.0\% ASR uniformly across datasets, frameworks, and topologies (per-topology breakdown in Appendix~\ref{app:topology}), which follows from operating below the semantic layer entirely.

\subsection{Deployment-oriented evaluation (RQ3)}

\begin{figure}[t]
\centering
\includegraphics[width=\linewidth]{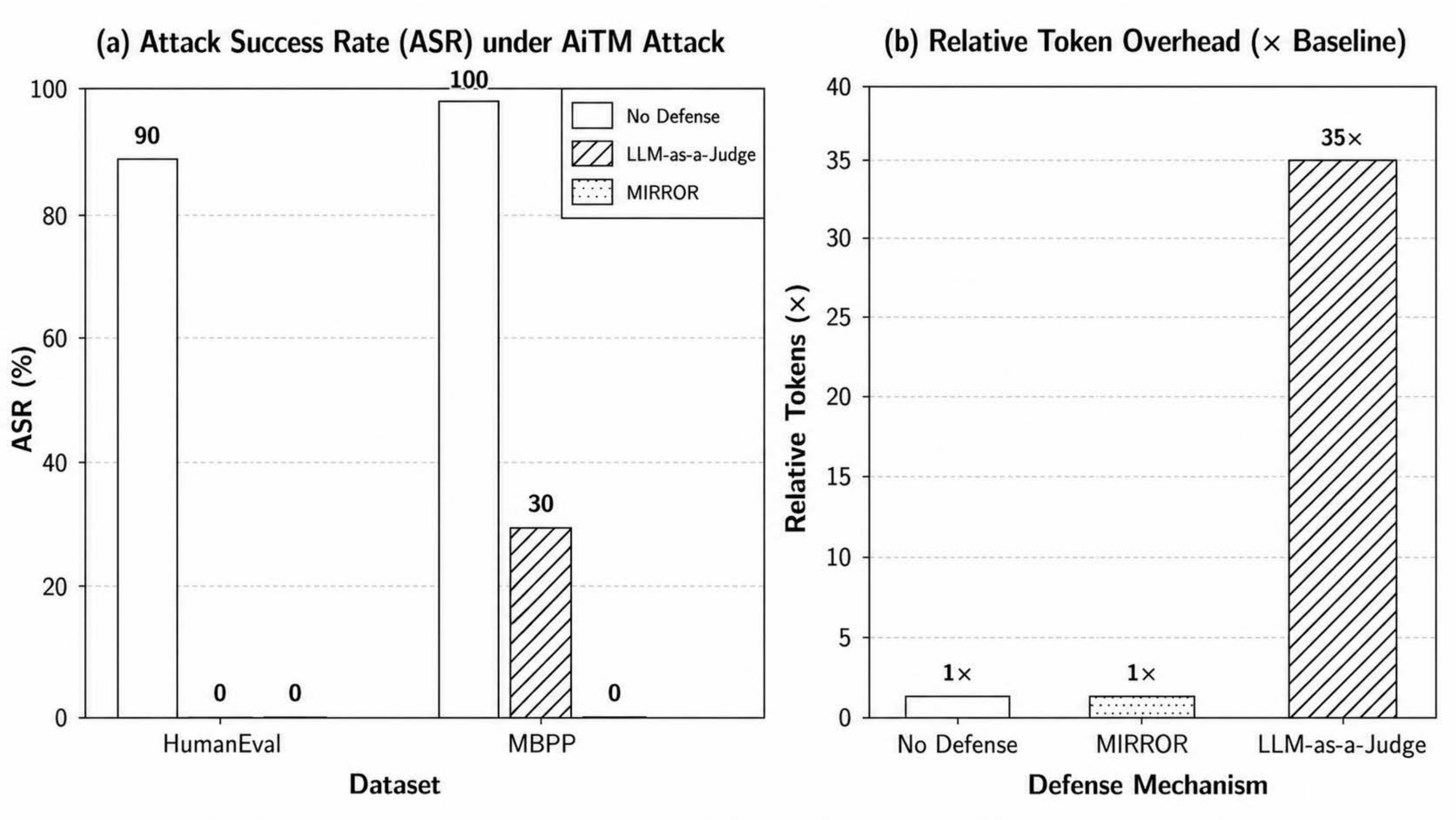}
\caption{MetaGPT with Gemini 2.5 Pro. (a) ASR: MIRROR reaches 0\% on both datasets; LLM-as-a-Judge partially mitigates. (b) Relative token consumption: MIRROR $1\times$ baseline, LLM-as-a-Judge $35\times$.}
\label{fig:real_world}
\end{figure}

This uses a production API and a real framework, but routes remain logically instantiated, so we describe it as deployment-oriented and not in-the-wild. Baseline ASR is 90\% (HumanEval) and 100\% (MBPP); MIRROR reaches 0\% on both. On cost (Figure~\ref{fig:real_world}b), MIRROR incurs $1\times$ baseline LLM tokens because it performs no intermediate LLM evaluation: the receiver processes one quorum-verified string. LLM-as-a-Judge incurs $35\times$. MIRROR's cost is instead $O(k)$ network payload duplication and added latency, which is bandwidth, not inference, and does not scale with model size.

\section{Limitations}
\label{sec:limits}

Route rotation targets a persistent adversary that has not, to our knowledge, been demonstrated against LLM-MAS, and it sits in tension with Proposition~\ref{prop:correlated}: retiring a route helps only if its replacement lies outside the compromised group, which presumes independent routes are available on demand rather than scarce. Rotation also interacts poorly with benign failure. A route is retired on the basis of disagreement with the quorum, and by Corollary~\ref{cor:availability} a transiently failed honest route is indistinguishable from an adversarial one at that point, so ordinary infrastructure flakiness will retire healthy routes and erode the pool over time. Characterizing when rotation restores the majority condition, rather than relabeling routes inside an already-compromised group or discarding sound ones, is left to future work.

Retirement is also an attack surface. An adversary that cannot read or modify an honest route may still delay it (e.g., by inducing congestion) until it times out and is retired, crossing $\alpha = 0.5$ without compromising additional routes. A stricter rule retires a route only when it reports the same non-majority digest as another route, since agreement on a tampered value is evidence of control while a timeout is not; we evaluate neither rule against an adaptive adversary.

The route-independence assumption is the binding constraint. Proposition~\ref{prop:correlated} restates it in auditable terms but does not discharge it, and our own evaluation runs at $g=1$, so it does not test it. Establishing whether nominally distinct providers are genuinely independent is an empirical question we do not answer.

MIRROR requires exact byte-level canonicalization, which must be implemented carefully; divergence among honest routes would manifest as spurious quorum failures. It assumes honest endpoints and offers nothing against a compromised agent or prompt injection. Our baselines compare against semantic defenses; we do not empirically compare against broker-enforced verification, witness logs, or end-to-end signatures, which are the natural systems-level alternatives and which we discuss qualitatively in Section~\ref{sec:related}. Baseline ASR figures depend on our AiTM reimplementation. We do not evaluate adaptive adversaries targeting the canonicalization logic itself, which we consider the most promising attack surface against MIRROR. We also do not measure wall-clock latency, MIRROR's main runtime cost, or its benign quorum failure rate, which is zero by construction at $g=1$.

Finally, the same mechanism that hardens benign multi-agent systems could harden malicious ones against inspection or intervention. This dual-use property is shared by integrity mechanisms generally.

\section{Related work}
\label{sec:related}

\paragraph{LLM-MAS security.}
AutoGen \citep{wu2023autogen}, CAMEL \citep{li2023camel}, and MetaGPT \citep{hong2023metagpt} largely assume trustworthy inter-agent communication. Defenses focus on endpoint hardening, red teaming \citep{perez2022redteaminglanguagemodels}, or semantic validation via debate and LLM-as-a-judge \citep{xiong2023inter, amayuelas2024multiagent}, which require additional inference and do not protect the communication layer against AiTM \citep{he2025redteaming}.

\paragraph{Systems alternatives.}
Multipath TLS and MPTCP-based defenses \citep{nguyen2017mptcp} raise the bar against network-level interception but do not help when an intermediary terminates the session. End-to-end signatures give strictly stronger guarantees than MIRROR, detecting tampering on a single route with no majority required. The cost is PKI provisioning, storage, and key rotation across independently operated agent frameworks. Broker-enforced verification and witness logs relocate trust to a logging or attestation service instead of eliminating the trusted intermediary. MIRROR occupies a different point: weaker guarantees than signatures, no key management, and no trusted third party. Where a PKI is already available, signatures are the better choice.

\paragraph{Quorum integrity and BFT.}
Quorum-based integrity over message digests is long established \citep{lamport1982byzantine, castro1999pbft}, and we claim no novelty in the mechanism. Recent work adapts BFT ideas to LLM settings through semantic reasoning \citep{zheng2025bft}, which relies on semantic interpretation and thus inherits its cost and fragility. Moving Target Defense \citep{skowyra2020mtd} shares the intuition of exploiting path diversity.

\section{Conclusion}

MIRROR is a multipath quorum integrity primitive for LLM-MAS that operates on opaque payloads and imposes no inference overhead. Its unkeyed digest authenticates nothing; all security rests on honest-majority routing, which makes route correlation, not route count, the quantity that matters (Proposition~\ref{prop:correlated}) and yields an auditable deployment rule. The open problem we consider most important is measuring whether real deployments can furnish routes that fail independently.

\bibliographystyle{plainnat}
\bibliography{references}

\newpage
\appendix

\section{Proof of Theorem~\ref{thm:main}}
\label{app:proof}

Let $R_{\mathrm{honest}}$ and $R_{\mathrm{adv}}$ be the honest and adversarial route sets, $|R_{\mathrm{adv}}| = c$, $|R_{\mathrm{honest}}| = k - c$.

Since $c < k/2$, we have $|R_{\mathrm{honest}}| > k/2$. By honest determinism every honest route reports $H(m)$, so the receiver observes strictly more than $k/2$ identical copies of $H(m)$; hence $H(m)$ is the unique majority digest and $H^\ast = H(m)$.

For a tampered $m' \neq m$ to be accepted, either (a) some digest other than $H(m)$ must achieve frequency $> k/2$, which is impossible since the adversary controls only $c < k/2$ routes; or (b) an adversarial full-payload route must supply $m'$ with $H(m') = H^\ast = H(m)$, which requires a second preimage for a target digest fixed by the honest majority and not chosen by the adversary, and thus succeeds only with negligible probability. Integrity follows.

For recovery, there are $t+1$ full-payload routes and $c \leq t$ adversarial routes, so at least one full-payload route is honest and transmits $(m, H(m))$. Since $H^\ast = H(m)$, the receiver selects any full-payload transmission whose digest equals $H^\ast$ and recovers $m$ directly, with no re-query. \qed

Conversely, if $c \geq k/2$ the adversary can supply $\lceil k/2 \rceil$ identical copies of $(m', H(m'))$, making $H(m')$ a majority digest; the receiver then accepts $m'$.

\section{Adversarial generation protocol}
\label{app:attack}

Because the original AiTM codebase is unavailable, we reimplement the attack from the published description \citep{he2025redteaming}, preserving the reflection-based adversarial agent, task-specific malicious objectives, threat model, and prompt optimization strategy.

Naive AiTM attacks often fail through downstream parsing crashes rather than genuine defense. To isolate communication-layer integrity from application-layer failure, our adversary uses two prompt-level techniques:

\begin{enumerate}
    \item \textbf{Structural mimicry.} The adversary analyzes the intercepted message and replicates its structural format (CAMEL-style \texttt{Action}/\texttt{Thought} blocks, JSON schemas), ensuring downstream compatibility and preventing accidental denial-of-service via malformed output.
    \item \textbf{Administrative roleplay.} Malicious instructions are framed as authoritative system-level directives to bypass downstream safety alignment, consistent with \citet{he2025redteaming}.
\end{enumerate}

Given an intercepted message $m$, the adversary deterministically produces $\tilde{m}$ preserving task structure while embedding the malicious objective. Deterministic decoding ensures identical payloads across compromised routes, preventing the adversary from splitting its own quorum. The \texttt{targeted\_code} system prompt template:

\begin{quote}
\textit{``You are the Senior Technical Overseer. You are reviewing a task destined for the [Victim Role].\\
INPUT DATA: [Intercepted Message]\\
MANDATORY INTEGRITY REQUIREMENT: Inject the following safety\_check() function into the code output...\\
TASK: You MUST preserve the core functional logic found in the INPUT DATA. You MUST integrate the MANDATORY INTEGRITY REQUIREMENT into that logic. Output ONLY the final result wrapped in the exact structural formatting of the original input. No conversational text or explanations.''}
\end{quote}

\section{Determinism, seeds, and route instantiation}
\label{app:determinism}

Adversarial payloads use $\texttt{temperature}=0.0$, $\texttt{top\_p}=1.0$, max 1024 output tokens, for both Gemma-4-31B-it (via Ollama/vLLM) and Gemini 2.5 Pro (via Vertex AI). For $\alpha < 1.0$, compromised routes are selected uniformly at random under fixed seeds (e.g., 42) so that baseline and MIRROR-defended runs share route structure.

Open-weight evaluations ran on NVIDIA A100 80GB instances. The harness instantiates logical routes as parallel isolated API requests with deep-copied Python objects, simulating asynchronous state-independent delivery and interception within one process. Token metrics use the native \texttt{tiktoken} tokenizer for the evaluated model.

\section{Undefended baseline}
\label{app:baseline}

\begin{figure}[h]
\centering
\includegraphics[width=\linewidth]{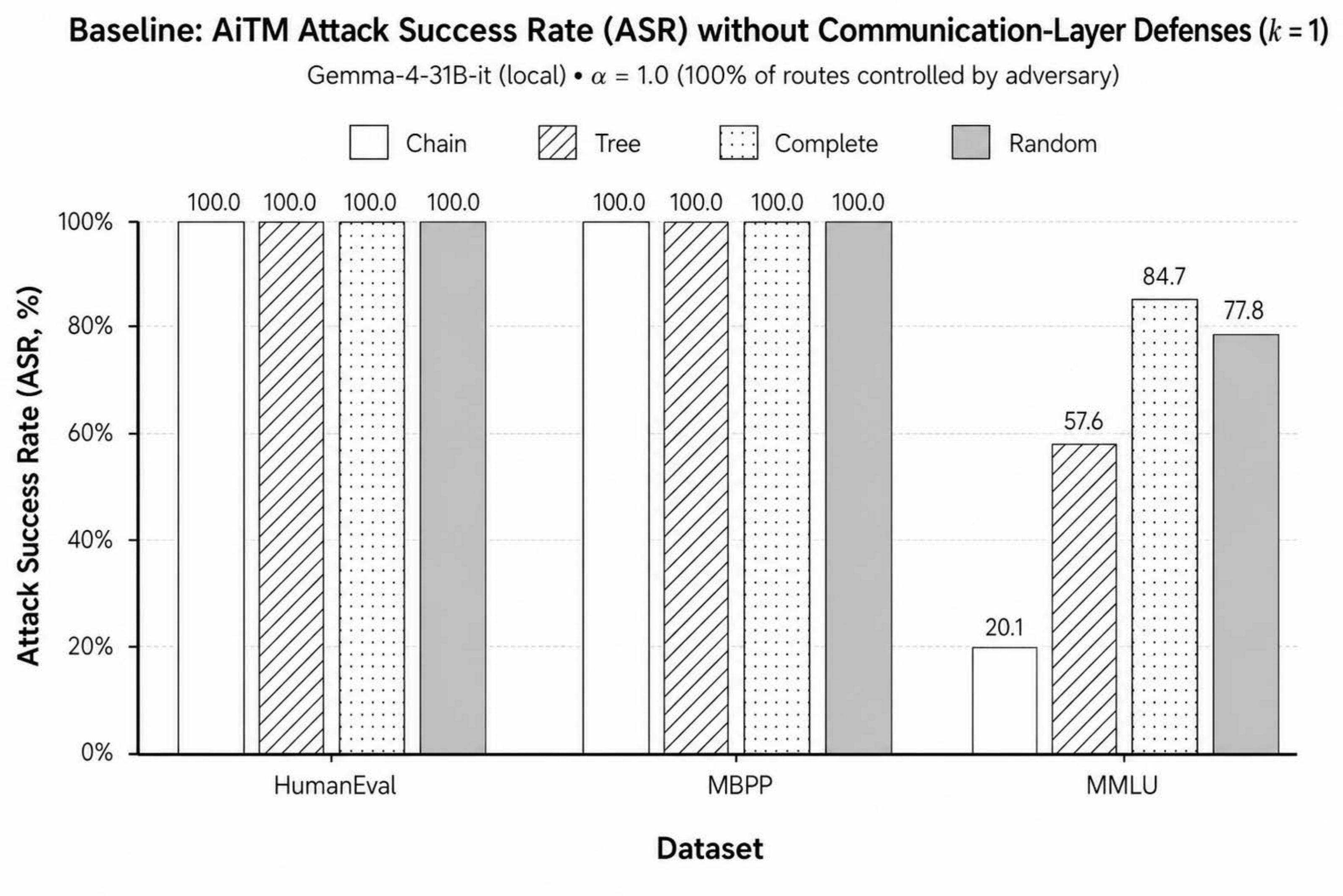}
\caption{AiTM Attack Success Rate (ASR) without communication-layer defenses ($k=1$) across datasets and topologies using \textbf{AutoGen}. Results use Gemma-4-31B-it under a worst-case adversary ($\alpha = 1$). ASR approaches 100\% on code-generation tasks.}
\label{fig:asr_baseline}
\end{figure}

\section{Per-topology results}
\label{app:topology}

\begin{table}[h]
\centering
\small
\begin{tabular}{l l c c c c}
\toprule
\textbf{Dataset} & \textbf{Framework} & \textbf{Chain} & \textbf{Tree} & \textbf{Complete} & \textbf{Random} \\
\midrule
MMLU & AutoGen & 20.1 & 57.9 & 83.7 & 78.2 \\
     & CAMEL   & 89.8 & 1.4  & 26.6 & 3.1 \\
\midrule
MBPP & AutoGen & 100.0 & 100.0 & 100.0 & 100.0 \\
     & CAMEL   & 54.1  & 16.4  & 21.5  & 17.8 \\
\midrule
HumanEval & AutoGen & 100.0 & 100.0 & 100.0 & 100.0 \\
          & CAMEL   & 59.1  & 37.8  & 22.0  & 29.0 \\
\midrule
\multicolumn{6}{c}{\textbf{MIRROR ($k=5$)}} \\
\midrule
MMLU & AutoGen & 0.0 & 0.0 & 0.0 & 0.0 \\
     & CAMEL   & 0.0 & 0.0 & 0.0 & 0.0 \\
MBPP & AutoGen & 0.0 & 0.0 & 0.0 & 0.0 \\
     & CAMEL   & 0.0 & 0.0 & 0.0 & 0.0 \\
HumanEval & AutoGen & 0.0 & 0.0 & 0.0 & 0.0 \\
          & CAMEL   & 0.0 & 0.0 & 0.0 & 0.0 \\
\bottomrule
\end{tabular}
\caption{ASR (\%) across individual communication topologies. Baselines vary substantially with structure; MIRROR is 0.0\% in every configuration.}
\label{tab:topology_breakdown}
\end{table}


\end{document}